\documentclass[nofootinbib,floats,floatfix,eqsecnum,prd,aps]{revtex4}
\usepackage[T1]{fontenc}
\usepackage[utf8]{inputenc}
\inputencoding{utf8}
\usepackage[dvips]{graphicx}
\usepackage{hyperref}
\usepackage{dcolumn,epsfig,minitoc}
\usepackage{amssymb,amsmath}
\usepackage[usenames]{color}
\def\o1{$O_{1}$}
\def\o2{$O_{2}$}

\def\lb{\label}
\newcommand{\beq}{\begin{equation}}
\newcommand{\eeq}{\end{equation}}
\newcommand{\bea}{\begin{eqnarray}}
\newcommand{\eea}{\end{eqnarray}}

\begin{document}

\title{Black hole solutions of gravity theories with nonminimal 
coupling between matter and curvature }
\author{Orfeu Bertolami$^{1,2}$}
\affiliation{$^1$Departamento de F\'isica e Astronomia, Faculdade de
Ci\^encias, Universidade do Porto, Rua do Campo Alegre s/n, 4169-007
Porto, Portugal \\$^2$Centro de F\'isica do Porto, Rua do Campo
Alegre s/n, 4169-007 Porto, Portugal}
\author{Mariano Cadoni$^{3,4}$ and  Andrea Porru$^{3}$}
\affiliation{$^{3}$Dipartimento di Fisica, Universit\`a di 
Cagliari.\\ $^{4}$INFN, Sezione di
Cagliari.\\ Cittadella Universitaria, 09042 Monserrato, Italy.}

\date{\today }

\begin{abstract}

We study black hole solutions in an extension of General Relativity 
(GR)
with an explicit non-minimal coupling between matter and curvature.
General black hole solutions satisfying the known energy conditions
are derived including the ones with  anti-de Sitter
background. These  solutions differ from those of 
GR just by a coupling function dependent rescaling of  
the mass and charge  of the black hole and by a ``dressing'' of the 
cosmological constant.
The existence of black hole solutions of the nonminimally 
coupled theory as well as the
conditions for a suitable weak field limit are considered as a
constraint on the coupling function responsible for the 
nonminimal coupling between matter and curvature. 
The ``dressing'' of the cosmological constant is then used to 
address the cosmological constant problem.

\end{abstract}

\maketitle
\tableofcontents

\section{Introduction}

As far as it is known General Relativity (GR) describes accurately all
gravitational phenomena at the solar system level \cite{GR0, GR} and
it gives rise to striking predictions among which black hole
solutions are possibly the most bizarre and surprising ones.
However, at galactic and cosmological scales, in order to account for
the observations two unknown components must be introduced: dark
matter, to explain the rotation curve of spiral galaxies and the
dynamical mass of clusters; and dark energy, an exotic form smooth
distribution of energy to account the late-time accelerated expansion
of the Universe. Strikingly, these two dark components constitute
about $95 \%$ of the energy content of the Universe and their nature
remains still a mystery.

Of course, several alternative gravitational theories have been
proposed to account for the observations usually explained by the
presence of dark matter and dark energy, such as, for instance,
$F\left(\mathsf{R}\right)$ theories of gravity \cite{fr1, fr2,
fr3}.
Another interesting alternative admits a non-minimal coupling between
curvature and matter \cite{modelo} (see Refs. \cite{cosm1, cosm2} for
proposals in the context of cosmology). The later is 
characterized by a coupling function $f(R)$ between matter and the 
curvature and  has a wide range of
theoretical and observational implications. It has bearings on
issues such as stellar stability \cite{stellar,stellar1}, preheating
after
inflation \cite{inflation}, mimicking of dark matter in galaxies
\cite{DM mim1} and clusters \cite{DM mim2} and the large scale effect
of dark energy \cite{DE mim1}. The issue of energy density
perturbations in the context  of an homogeneous and isotropic
cosmological background was studied in Ref. \cite{frazao}. For a
complete and updated review of the main developments involving this
type of gravity model with nonminimal coupling between matter and
curvature, the reader is referred to Ref. \cite{review}.

In this work, we examine static, spherically symmetric black hole
solutions in the context  of  theories with non-minimal coupling
between matter and curvature. The existence and the properties 
of these solutions is an important  test for all alternative 
theories of gravity. Of particular relevance is the deviation of the 
black solution from the usual ones of GR, which is expected to
provide  
useful information about the strong coupling regime of the theory.

As we shall see, the  study of black hole solutions of non-minimally 
coupled (NMC) gravity theories  has a rich
set of implications once  the presence of various  matter sources 
is considered (so
to test the coupling term) and once non-trivial backgrounds are
admitted. As expected, a relevant matter when studying these
solutions is whether they saturate or not the null-energy condition
(NEC). The solutions that saturate the NEC, which will be 
discussed at length in this paper, represent the 
generalization to the non-minimally coupled theory of the well-known 
Schwarzschild and Reissner-Nordstrom  solutions (both in flat and 
anti de Sitter space) of GR.

Owing to the non-minimal coupling  between 
matter sources and curvature, one naively  expects strong deviation 
of these solutions from  their  behaviour in GR. Most particularly in
what 
concerns their uniqueness. As we shall see, our 
investigation does not confirm this naive expectation.  The 
generalization of the Schwarzschild and Reissner-Nordstrom  solutions 
are very similar to their GR counterparts (they differ just by a 
coupling function dependent rescaling of their mass and charge and by 
a dressing of the cosmological constant). 
Moreover, we will find strong evidence that they are also essentially 
unique.  
 
Another important point when dealing with  gravity theories  
with with non-minimal coupling
between matter and curvature is the issue of the determination of the 
coupling function $f(R)$. The NMC theory is regarded 
as an effective theory with the coupling function $f(R)$ 
parametrizing our ignorance  about a presumed  fundamental 
theory of gravity. On the other hand, the local behaviour of the 
coupling function $f(R)$ encodes the physical information on the 
behaviour of the NMC theory at length scales $\sim R^{-1/2}$.
Within this phenomenological perspective it is 
very important to gather information coming from different length 
scales  and coupling regimes of gravity, so to constrain
the analytic form of $f(R)$.

Until now most of the constraints on the coupling function 
came from the large scale, infrared (IR), behaviour of gravity
\cite{DM 
mim1,DM mim2,DE mim1,review}. On its hand, the investigation on 
black holes is expected to provide insight not only about the IR 
behaviour of $f(R)$ (through the  weak field limit) but, owing to the 
presence of singularities, also on its ultraviolet (UV) behaviour.
We shall show that this is the case. Investigation of the black hole 
solutions  will allow us to establish constraints on the  local
behaviour of 
$f(R)$ at the black hole singularity and at the flat asymptotic 
region. As a byproduct of our research  we consider the dressing of 
the cosmological constant to examine the possibility to generate a
natural hierarchy of mass 
scales between the bare and the dressed cosmological constant.

This work is organised as follows. In the next section we review the
main features of the non-minimally coupled matter-curvature model
\cite{modelo} that we have in mind. In section \ref{SIII}, we examine
the
energy conditions suitable to study black hole type solutions, most
particularly NEC. We analyze then static, spherically symmetric
solutions of the alternative theory of gravity under the conditions
that the trace of the energy momentum tensor is constant and that the
components satisfy the condition $T_{0}^{0}=T_{1}^{1}$ (sections 
\ref{SIV} 
and \ref{sectionbh}). How natural
are these coditions will be then discussed. Next, de Sitter and
anti-de
Sitter backgrounds are considered once a cosmological constant is
introduced  (section \ref{SVI}). Solutions in these backgrounds are
studied
in section \ref{SVII}. In section \ref{SVIII}, charged black hole
like solutions are sought in
various backgrounds and under the conditions set by the NEC. In the
remaining sections we study the weak field limit (section \ref{wf}),
the
constraints that
the existence of the considered solutions and the weak field limit
pose on the coupling function (section \ref{cf}) and use the freedom
in
choosing the 
coupling function to address the cosmological
constant problem (section \ref{acc}). Finally, section \ref{SXII}
contains our
conclusions.

\section{Field equations and stress energy tensor}
\lb{SII}
In this  paper we consider  a gravity theory in which the coupling 
between the gravitational field and matter is non-minimal. 
Invariance of the action  under diffeomorphisms together with the 
requirement that in the flat limit the Lagrangian for the matter 
fields  $\mathcal{L}_m$  reduces to that in Minkowski space, fixes 
the dependence of $\mathcal{L}_m$  on the metric tensor $g_{\mu \nu}$.
On the other hand, these requirements leave open the possibility that 
$\mathcal{L}_m$  enters in the action not in the minimal way 
(multiplied by the covariant volume element) but nonminimally, trough 
an arbitrary  function of the scalar curvature. 

We are therefore lead to  consider for the matter-gravity coupled 
system an  action of the form \cite{modelo}:

\begin{equation}\label{S1}
S = \int \left \{ \frac{1}{2} R +[1+ \lambda
f(R)] \mathcal{L}_m
\right \} \sqrt{- g} d^4 x,
\end{equation}
where $f(R)$ is an arbitrary function of the curvature scalar,
we have chosen units such that $8\pi G=1$ and $\lambda$ is a 
constant parameter. As mentioned in the introduction, the gravity
theory described by the action (\ref{S1}) has 
been investigated given its interesting cosmological implications and
the possibility for a gravitational solution for the 
dark matter \cite{DM mim1,DM mim2} problem and the large scale effect
of dark energy \cite{DE mim1}.

Varying the action with
respect to the metric $g^{\mu \nu}$, we get the field equations for 
the gravity field, 
\begin{equation}\label{feq}
G_{\mu \nu} = - 2\lambda f_R
\mathcal{L}_m R_{\mu \nu} + 2 \lambda (\nabla _\mu \nabla _\nu -
g_{\mu \nu} \Box)  \mathcal{L}_m f_R + [1 + \lambda f(R)] T_{\mu \nu}
,
\end{equation}
where $G_{\mu \nu}$ is the Einstein's tensor, $f_R(R) = \frac{d f}{d
R}$ and  $T_{\mu \nu}$ is the
energy-momentum tensor,
\begin{equation}
T_{\mu \nu} =- \frac{2}{ \sqrt{- g} } \frac{\delta \sqrt{- g}
\mathcal{L}_m }{ \delta{g^{\mu \nu}}}.
\end{equation}

By contracting Eq. (\ref{feq}) with $g^{\mu \nu}$ we get  the general
form
for the scalar curvature:
\begin{equation}\label{R}
R= \frac{- 6 \Box (\lambda \mathcal{L}_m f_R)+ [1 + \lambda f(R)] T}{2
\lambda \mathcal{L}_mf_R - 1}.
\end{equation}
Thus, due to the nonminimal coupling  between scalar curvature
and 
matter, $R$ is no longer proportional to the trace of the
energy-momentum tensor, but it also depends on the higher order
derivative of the coupling term.

Also the field equations for the matter fields, obtained by varying 
the action (\ref{S1}) with respect to the matter fields  will get a
dependence 
from  the scalar curvature  an its derivatives.  In this paper we 
are mainly  concerned in the dynamics of gravity. In the following
we will, therefore,  
use the field equations for matter fields only when needed to solve 
the dynamics of the gravitational  field.

In the following we will need an expression relating 
 the higher order derivative coupling term
and the first two components of the energy-momentum tensor.
Some algebraic manipulation renders from the field equations 
(\ref{feq})
\begin{equation}\label{mattercon1}
\frac{[1 + 2 \lambda f_R \mathcal{L}_m]}{[1+
\lambda f]}(R_{0}^{0}-R_{1}^{1}) + \frac{[   
\nabla_1 \nabla^{1}- \nabla_0 \nabla^{0}] (2 \lambda \mathcal{L}_m
f_R)}{[1+
\lambda f]} = T_{0}^{0}-T_{1}^{1}.
\end{equation}

The nonminimal nature of the gravity-matter coupling is fully evident 
in the RHS of Eq. (\ref{feq}). Notice that the curvature appears also 
as  source of the gravitational field so that $T_{\mu\nu}$ is not 
covariantly conserved as in GR  but satisfies 
the equation
\begin{equation}\label{e1}
\nabla ^{\mu} T_{\mu \nu} = \frac{\lambda f_R}{ 1 + \lambda f}[g_{\mu
\nu} \mathcal{L}_m - T_{\mu \nu}] \nabla^{\mu} R,
\end{equation}
as one can easily verify  taking into account the contracted  
Bianchi identities, $\nabla^{\mu}G_{\mu \nu} =0$
and the following identity:
\begin{equation}
(\Box \nabla_\nu - \nabla_\nu \Box)f_R = R_{\mu \nu} \nabla ^\mu f_R.
\end{equation}

Physically, equation (\ref{e1}) describes an exchange of energy and
momentum between
 matter and  higher order derivative curvature terms.
An important point to consider are the conditions to
be   
satisfied in order that 
this 
exchange of energy and momentum does not take place. One can easily 
see that  the 
covariant derivative of $T_{\mu\nu}$, turns to zero only in two cases 
(apart  
from the obvious case $f(R)=const.$, corresponding to GR).
\begin{enumerate}\lb{w1ss}
 \item
$T_{\mu\nu}=g_{\mu\nu}\mathcal{L}_m$ \\
\item  
$R=const.$ 
\end{enumerate}
In the first case, $\,\mathcal{L}_m$ does not depend explicitly on
the 
metric.
This  corresponds to
terms in the matter Lagrangian in which the 
coupling 
to gravity is only through the covariant volume (e.g a term
$\sqrt{-g} \,V(\phi)$ 
for a scalar field or a cosmological constant).
The most interesting physical case belonging to this class is 
represented  by  a cosmological constant,
\begin{equation}\lb{b1}
\mathcal{L}_m= -\Lambda, 
\end{equation}
which in GR gives origin to maximally symmetric spaces: Minkowski,
de 
Sitter (dS) and anti-de Sitter (AdS), corresponding respectively to
vanishing, 
positive and negative $\Lambda$.

The second case also admits as solutions  the  maximally symmetric 
spaces. The other interesting cases belonging to this class are those
spaces 
with constant scalar curvature, which are not maximally symmetric.
In  GR they describe 
vacuum solutions ($T_{\mu\nu}=0$) like  the Schwarzschild (SCHW)
solution  or,
conformal invariant matter ($T=T_{\mu}^{\mu}=0$)
like the Reissner-Nordstrom (RN) solution.

The extension of the above-mentioned  solutions of GR to the case of
nonminimally coupled 
gravity theories is the main goal of this paper. 
Here, we just point out that this extension is far from being trivial
and 
in general it requires non trivial constraints on the form  of 
the coupling function $f(R)$.
This is quite evident from the form of the field equations
(\ref{feq}) 
and from the expression of the scalar curvature (\ref{R}).
Differently from GR, in the RHS of these equations enters 
the  scalar curvature,  through  the coupling function $f(R)$. For 
instance, we see from equation (\ref{R}) that a traceless  
matter stress-energy tensor does not automatically imply vanishing of 
the scalar curvature of the space-time. Additional constraints on the 
form   of the coupling function $f(R)$ (or on the form of 
$\mathcal{L}_m$) are needed in order to achieve that.

For the same reason, the question whether the  existence of other  
solutions, apart from the GR solutions, is clearly nontrivial.
 One could consider for instance
solutions with 
scalar hair. However, in this paper  we will limit our 
investigation to the extension  of the  well-known GR solutions 
(maximally symmetric spaces, the Schwarzschild  and  
Reissner-Nordstrom    solutions, both in Minkowski and AdS space).  

\section{Energy conditions}
\lb{SIII}
Energy conditions in GR, and in particular the  the null-energy
condition 
(NEC) and the  strong-energy
condition (SEC),    correspond to a fairly general way to translate 
geometric  informations on the congruence of geodesics into  
conditions for the matter stress energy tensor.
In the context of a nonminimally coupled gravity theory, one
naturally 
expects the energy  conditions to be drastically modified by the
presence of  
the curvature dependent terms in the RHS of Eq. (\ref{feq}).
For the derivation of the energy conditions in nonminimally coupled 
theory we follow closely Ref. \cite{OBSEqueira}.

In order to derive  the NEC and the SEC, one usually considers 
the Raychaudhuri equation
together with the request that  gravity is attractive.
In the case of a congruence of a time-like geodesic defined by a
vector field $u^\mu$ this equation reads:
\begin{equation}\lb{tlg}
\frac{d \theta }{ d \tau} = - \frac{1}{3}\theta^2 - \sigma_{\mu \nu}
\sigma^{\mu \nu} +\omega_{\mu \nu} \omega^{\mu \nu}- R_{\mu \nu}
u^\mu u^\nu,
\end{equation} 
where $\theta$, $\sigma_{\mu \nu}$, $\omega_{\mu \nu}$ 
are, respectively, the
expansion parameter, the shear and the rotation associated to the
congruence.

For  null geodesics $k^\nu$,   the Raychaudhuri equation is
given by:
\begin{equation}\lb{nlg}
\frac{d \theta }{ d \tau} = - \frac{1}{2} \theta^2- \sigma_{\mu \nu}
\sigma^{\mu \nu} +\omega_{\mu \nu} \omega^{\mu \nu}- R_{\mu \nu}
k^\mu k^\nu.
\end{equation} 
These equations are purely geometric and independent of the gravity
theory we are considering. The connection with the gravity theory
arises when we express Eqs. (\ref{tlg}) and (\ref{nlg}) in terms of
$T_{\mu \nu}$.
Assuming that gravity is attractive  (convergence of geodesics)
we have $\frac{d
\theta}{ d \tau} < 0$ and , since $\sigma_{\mu \nu} \sigma^{\mu \nu}
\geq 0$, for any hypersurfaces of orthogonal congruences (for
vanishing 
rotation), we get the
following conditions:
\begin{itemize}
\item \textit{Strong-energy condition (SEC)}
\begin{equation}\label{sec}
R_{\mu \nu} u^\mu u^\nu \geq 0;
\end{equation}
\item \textit{Null-energy condition(NEC)}
\begin{equation} \label{nec}
R_{\mu \nu} k^\mu k^\nu \geq 0.
\end{equation}
\end{itemize}

In the GR case, one then uses  Einstein's field equations into the NEC
and SEC 
conditions (\ref{nec}) and (\ref{sec}).  We get respectively:
\begin{equation}
T_{\mu \nu }k^\mu k^\nu \geq 0; \quad \left ( T_{\mu \nu } - g_{\mu
\nu} \frac{T}{2} \right) u^\mu u^\nu  \geq 0.
\end{equation}

On the other hand, in the case of the nonminimally coupled theory 
(\ref{S1}) we have to introduce an effective coupling constant 
$\kappa$ and  an additional energy momentum tensor $\hat T_{\mu 
\nu}$ describing the effects of the nonminimal coupling between 
matter and gravity. 

We start by  writing Eqs. (\ref{feq}) in the form:
\begin{equation}\lb{d1}
G_{\mu \nu} = \kappa (\hat T_{\mu \nu} + T _{\mu \nu}),
\end{equation}
where
\begin{equation}\label{ecc}
\kappa= \frac{1 + \lambda f(R)}{1 +2 { \mathcal{L}_m}\lambda f_{R}
(R)},
\end{equation}
which is an effective coupling constant, and
\begin{equation}\lb{est}
 \hat T_{\mu \nu}=\frac{1}{[ 1+ \lambda f]}\Bigl \{
-\lambda \mathcal{L}_m f_{R}R g_{\mu \nu}  +(\nabla_\mu
\nabla_\nu - g_{\mu \nu}
\Box)(2\lambda \mathcal{L}_m f_{R}) \Bigr \}.
\end{equation}

In order to keep gravity attractive, we have the additional condition
$\kappa >0$.\\

If we define $\hat T $ as the trace of $\hat T_{\mu \nu}$, using
Eqs. (\ref{d1}) into Eqs. (\ref{sec}) and (\ref{nec}), one obtains:
\begin{itemize}
\item \textit{SEC condition}
\begin{equation}
R_{\mu \nu} u^\mu u^\nu =\left [ \kappa ( \hat T_{\mu \nu} + T_{\mu
\nu})
+\frac{1}{2} g_{\mu \nu}\bigl ( \hat T + T\bigr) \right ]u^{\mu}
u^\nu \geq 0;
\end{equation}
\item \textit{NEC condition}
\begin{equation}
R_{\mu \nu} k^\mu k^\nu =\left [ \kappa (\hat T_{\mu \nu} + T_{\mu
\nu})
+\frac{1}{2}g_{\mu \nu} \bigl ( \hat T + T\bigr) \right ]k^\mu k^\nu
=\kappa  (\hat T_{\mu \nu} + T_{\mu \nu}) k^\mu k^\nu \geq 0.
\end{equation}
\end{itemize}

By choosing a suitable reference frame, one can always take  
a  null vector of  the form
\begin{equation}
k^\mu = \left (
\begin{array}{lccr}
-1 & 1 & 0 & 0
\end{array}
\right ).
\end{equation}
Assuming $\kappa >0$ the NEC conditions become
\begin{equation}\lb{l1}
-( \hat T_{0}^{0} +T_{0}^{0})+ (\hat T_{1}^{1}+ T_{1}^{1}) \geq 0.
\end{equation}
We can write this condition in the form:
\begin{equation}\label{NEC}
T_{0}^{0}- T_{1}^{1} \leq \frac{2\lambda}{(1+\lambda f)} \bigl [(
\nabla_1 \nabla^{1} -  \nabla_0 \nabla^{0}) (\mathcal L_ m f_R) \bigr
].
\end{equation}
The latter represents  a generalization of  the usual NEC of GR to
nonminimally 
coupled theories. \\

\subsection{NEC saturation}
 
In GR, NEC saturation sets a strong constraint on the form of the
matter sources.
In fact, for GR saturation of the inequality (\ref{l1}) implies
\begin{equation}\label{NECgr}
T_{0}^{0}= T_{1}^{1}.
\end{equation}
The previous relation is satisfied, in particular,   by
matter sources 
of particular interest, on which this paper is focused: $(a)$ 
The vacuum , $(b)$ Electromagnetic charged sources and  (c) 
Cosmological constant.

Upon using Einstein equations, Eq.  (\ref{NECgr}) translates into a
condition 
for the first two components of the Ricci tensor,
\beq\lb{s2}
R_{0}^{0}=R_{1}^{1}.
\eeq

We can reformulate the previous statement in a different way.  In GR,
Eqs. (\ref{NECgr}) 
and (\ref{s2}) allow two 
independent ways to implement the saturation of the NEC: 
in terms of sources alone or in terms of the gravitational field
alone 
(i.e. in the space of the solutions) and if we 
impose them  simultaneously the field equation (\ref{mattercon1})
becomes 
an identity.

The situation is somehow  different in the case of  the non minimally
coupled theory.
In this latter case, 
NEC saturation leads to the following condition:
\begin{equation}\label{NECnm}
\frac{[-\nabla_{0}^{0} +\nabla_{1}^{1}] 
(2\lambda f_R \mathcal{L}_m)}{1+\lambda f} = T_{0}^{0}-T_{1}^{1}. 
\end{equation}

Thus, in nonminimally coupled theories, NEC saturation is not a
simple
constraint on the form of  the matter sources, but it leads
to a relation between  matter source  and derivatives 
of the coupling function. Nonetheless, upon use of the fields
equations, NEC saturation  is equivalent to the same GR condition
(\ref{s2}).
In fact using  Eqs. (\ref{NEC})  and (\ref{mattercon1}),  one
immediately 
finds Eq. (\ref{s2}).

These features have an important consequence, which we will use 
later for deriving static, spherically solutions  of the nonminimally
coupled 
 theory. We cannot ``trivialize'' the field equation
(\ref{mattercon1}) by imposing the condition (\ref{s2})  and a
condition on 
the sources. Given the presence of the  term depending on  
 $f(R)$  in Eq. (\ref{NECnm}), an additional 
condition involving the coupling function $f$ is needed to transform 
Eq. (\ref{mattercon1}) into an identity.

This also implies that, analogously to GR,  in  the nonminimally 
coupled theory,
once we impose the condition (\ref{s2}), the NEC 
conditions  (\ref{NECnm})  becomes  an identity by virtue of the  
field equations.

\section{Static, spherically symmetric solutions}
\lb{SIV}
We are interested in static, spherically symmetric solutions of the 
field equations (\ref{feq}).
We consider a  static spacetime with spherical symmetry and use 
the following 
parametrization for the metric:
\begin{equation}\label{ansatz}
ds^2= - e ^{2 \alpha} dt^2 + e^{-2 \alpha} dr^2 + H(r)^2 d \Omega^2,
\end{equation}
where $\alpha(r)$ and  $H(r)$ are functions of the radial coordinate. 
Setting  $H^2 =
e^{2 \rho}$, the non vanishing  components of the Ricci tensor are:

\begin{equation}\label{ricci}
\left \{
\begin{array}{ll}
R_{00} = [\alpha'' + 2 \alpha '^2 + 2 \rho ' \alpha'
]e^{4\alpha},\\
R_{11} = -[\alpha'' + 2 \alpha '^2 +2 \rho '' +2\rho'^2  + 2
\rho '
\alpha'   ],\\
R_{22} = -[\alpha' \rho' + 2 \rho '^2 + \alpha' \rho' + \rho''] e^{2
(\alpha+\rho)} +1,\\
R_{33} = \sin^2 \theta  R_{22}.
\end{array}
\right.
\end{equation}
where the prime denotes derivatives with respect to the radial 
coordinate $r$ and $i=0,1\ldots 3$ indicates respectively the 
coordinates $t,r,\theta,\varphi$.

One can now  easily write down   the field equation (\ref{feq}) for 
the static,  spherically symmetric case. We have three  equations for 
only two independent metric functions ($\alpha(r)$ and $H(r)$). The 
system is in general overconstrained unless we treat $f(R)$ as a 
dynamical variable, i.e determined by the field equations.
We will see later in detail that  generalization to nonminimally 
coupled theory of the most interesting solution of GR imposes very 
weakly constraints on the form of $f(R)$, typically they will just 
constrain its value at some points.

As independent equations we can obviously  choose linear combinations 
of the field Eqs. (\ref{feq}). In particular, we have
\begin{equation}\label{IIfe}
-R_{0}^{0} + R_{1}^{1} =-2 e^{2\alpha - \rho} \frac 
{d^2e^\rho}{dr^{2}},
\end{equation}
which used in Eq.  (\ref{mattercon1}) gives 
\begin{equation}\label{mattercon2}
\frac{2 e^{2\alpha}} {[1+
\lambda f]} \left[\left(1 + 2 \lambda f_R \mathcal{L}_m\right) e^{-
\rho } \frac 
{d^2e^\rho}{dr^{2}}+\lambda 
\frac{d^2}{dr^{2}} \left(\mathcal{L}_m f_R\right)\right] = T_{0}^{0}
- T_{1}^{1},
\end{equation}

Let us now write the NEC  (\ref{NEC}) and the NEC saturation relation 
(\ref{s2}) in the static, spherically symmetric case. We obtain 
respectively:
\begin{equation}\label{NEC1}
T_{0}^{0}-  T_{1}^{1} \leq \frac{2\lambda  e^{2\alpha}}{(1+\lambda
f)} \bigl[
\frac{d^2}{dr^{2}}(\mathcal L_ m f_R) \bigr],
\end{equation}
and
\begin{equation}\label{NECsat1}
\frac{d^2e^\rho }{dr^{2}}=0.
\end{equation}

Up to  irrelevant integration constants, we find from (\ref{NECsat1})
\begin{equation}
e^\rho = r.
\end{equation}

Our main goal is the generalization to the nonminimally
coupled
theory (\ref{S1}), of the static, 
spherically symmetric solution of GR, namely the AdS
spacetime, 
the SCHW and  the RN black hole solutions, both in flat and AdS space.
In oder to derive these solutions, we use the following strategy.
We will first write down the general condition that must be satisfied 
by the energy 
momentum tensor of the corresponding matter sources.  The next step
is the use of 
the NEC. All the above solutions satisfy the NEC in GR but,  in 
principle, they  allow for two kinds of generalizations in the
nonminimal 
theory: $(a)$  Solutions satisfying the NEC in the
non-minimally 
coupled theory; $(b)$ Solutions not satisfying the NEC in the
non-minimally 
coupled theory.  In general, these two classes of solutions have a 
complete different behaviour. The final step is  integrating  
the field equations and, when possible, writing down the solution in
an explicit form.

\section{Sources characterized by 
\texorpdfstring{$ T=constant$}{T=Constant} and  
\texorpdfstring{$T_{0}^{0}=T_{1}^{1}$}{T00=T11}}
\label{sectionbh}

The simplest matter sources  to  be considered  
in the static, spherically symmetric case are those  saturating the 
NEC in GR and characterized additionally by a constant trace of the 
energy momentum tensor, i.e
\begin{equation}\label{p1}
T_{0}^{0}=T_{1}^{1}, \quad T=constant= a  .
\end{equation}
One can easily realize that conditions (\ref{p1}) are, in particular, 
satisfied by  the Minkowski vacuum, by a cosmological constant and, 
away from the position of the source,  by
charged and 
uncharged pointlike 
sources of mass $M$.

Although the matter sources characterized by
Eqs. 
(\ref{p1})  always saturate  the NEC of GR  (\ref{NECgr}), the 
corresponding solutions  may  
or may not   saturate the NEC for the NMC theory (\ref{NEC1}).
Let us therefore discuss these two cases separately.

\subsection{NEC saturating solutions}
\lb{ff1}
In this case the solutions and their matter sources  must satisfy
simultaneously 
Eqs. (\ref{p1}), (\ref{NECsat1}) and saturate the NEC for non 
minimally coupled theories, Eq. (\ref{NEC1}). These solutions
saturate 
both the NEC of GR and those of the NMC theories.

Inserting Eq. (\ref{p1}) into 
Eq. (\ref{NEC1}), we get
the simple  condition:  
\begin{equation}\label{p10}
\frac{d^2}{dr^2}\left(\mathcal{L}_m f_R\right)=0.
\end{equation}
Eq. (\ref{p10}) can be easily solved to give,
\begin{equation}\label{chnulcon}
(\mathcal{L}_m f_R)= \xi r +k,
\end{equation}
where $\xi$ and $k$ are integration constants.\\

When $\xi\neq 0$ Eq. (\ref{chnulcon})   yields  a stringent 
constraint on the form of 
the coupling $f(R)$. On the other hand, for  $\xi= 0$, it renders
a very weak constraint on the form of $f(R)$. Let us discuss these
two 
cases separately. \\

\centerline{
\fbox{$\xi =0$}} 

\bigskip
In this case we have 
\begin{equation}\label{chnulcon1}
\mathcal{L}_m f_R= k.
\end{equation}

From Eq. (\ref{R}) we get:
\begin{equation}\label{r1}
R= \frac{[1+ \lambda f(R)] a}{2\lambda k-1}.
\end{equation}

This equation implies that the scalar curvature of the spacetime must
be
necessarily 
constant $R=R_{0}$. The value of $R_{0}$ is obtained by solving Eq. 
(\ref{r1}). 

An important consequence of having a spacetime of 
constant curvature is that the stress-energy tensor $T_{\mu \nu}$ is 
conserved. This follows immediately from 
Eq. (\ref{e1}).
We have now two possible cases to examine:\\

$(1)$ $\,\mathcal{L}_m$ is not constant\\
In this case the only way to 
solve Eq. (\ref{chnulcon1}) is to set,
\begin{equation}\label{k2}
f_R(R_{0})= k=0.\\
\end{equation}
and Eq. (\ref{r1}) gives
\begin{equation}\label{hny}
R_{0}= -[1+ \lambda f(R_{0})] a.\\
\end{equation}

$(2)$ $\,\mathcal{L}_m= constant=b$\\
In this case Eq. (\ref{chnulcon1}) gives 
\begin{equation}\label{k1}
f_R(R_{0})= \frac{k}{b}.   
\end{equation}
Similarly to the previous case,  the scalar curvature $R_{0}$ is
obtained by solving Eq. (\ref{r1}).

Since the scalar curvature is constant, in both cases $(1)$ and
$(2)$, from Eq. (\ref{e1})
it follows immediately that for the solution belonging to  this class,
the energy-momentum tensor is covariantly conserved.

Constancy of the scalar curvature together with Eq. 
(\ref{chnulcon1}), implies, apart from constant 
covariance of $T_{\mu\nu}$, a drastic simplification  of the 
field Eqs. (\ref{feq}):
\begin{equation}\label{feq1}
R_{\mu \nu} =\frac{1}{(1+2 \lambda k)} \left[(1+ \lambda f(R_{0}))
T_{\mu 
\nu}+ \frac{1}{2} g_{\mu\nu}R_{0}\right] . 
\end{equation}

By defining a rescaled stress-energy tensor and an ``effective'' 
cosmological constant,
\begin{equation}\label{eeqre}
\tilde {T}_{\mu \nu} =  \left [ \frac{1+ \lambda f(R_{0})}{1+ 2
\lambda 
k} \right ] T_{\mu \nu},\quad \tilde \Lambda= 
\frac{1}{2}\frac{R_{0}}{1+2\lambda k}.
\end{equation}
the previous equation takes the form of  Einstein field equations of 
general relativity with stress energy tensor $\tilde T_{\mu\nu}$ and
a cosmological constant 
$\tilde \Lambda$.

It is important to stress that the conditions for the existence of
this 
class of solutions sets very weak constraints on the form of the 
coupling function $f(R)$. In fact equation (\ref{k2}) 
determines the value of the derivative of $f$ in a point, whereas Eq. 
(\ref{k1}) does not impose any constraint on the form of $f(R)$ as  
the integration constant 
$k$ remains undetermined.\\

\centerline{
\fbox{$\xi \neq0$}} 
\bigskip

In opposition to the previous case, Eq. (\ref{chnulcon})  gives
stringent 
constraints on the form of $f(R)$. Together with the field equations 
(\ref{feq}) it gives a system of two independent differential 
equations for two variables $\alpha, f(R)$. Thus, in general the 
function $f(R)$ is completely determined, apart from integration 
constants, by the dynamics of the 
gravitational field .

Moreover, the function $f(R)$ is 
also constrained by the asymptotic behaviour of the scalar curvature 
$R$. One can easily show that  solutions  with flat or AdS 
asymptotics require a non-analytic behaviour of $f(R)$.

To show this fact,  we examine the $r\to\infty$ behaviour of the
fields and 
we assume that in this limit the curvature goes to a constant $R_{0}$
(with $R_{0}=0$ in the flat case). The asymptotic behaviour of 
$\mathcal{L}_m $ can be found by assuming that matter fields in the
action (\ref{S1})
have at most an  infrared divergence  due the 
infinite volume  of the spacetime.  Thus, at leading order 
in the $r\to \infty$, we have 
$\mathcal{L}_m= \mathcal{O}(1)$. With this, Eq. 
(\ref{chnulcon})  gives for $r\to\infty$
\beq\lb{da}
f(R)= \mathcal{O}(r).
\eeq
Assuming that $f(R)$ is analytic in $R_{0}$ we can expand it in power
series 
of $R-R_{0}$. Substituting  this into Eq. (\ref{da}) one easily 
finds inconsistency  with  the assumed analyticity of $f(R)$ in
$R_{0}$.

In the following sections we shall consider explicit black hole
solutions 
of our nonminimally coupled theory, 
for  the case of the simplest, and most 
interesting, matter sources  satisfying 
Eq. (\ref{p1}):\\
(a) Maximally symmetric spaces;\\
(b) Generalization of 
Schwarzschild black hole solution of GR 
($T_{\mu\nu}=0$ away from the source); \\
(c)  Charged black hole solutions in flat 
spacetime, the generalization 
of the Reissner-Nordstrom (RN) 
solution 
($T_{\mu\nu}\neq 0, T=0$); \\ 
(d) Schwarzschild black hole solutions
in 
AdS  spacetime (SAdS);\\
(e) RN solutions in 
AdS  spacetime (RNADS).\\
Both solutions (d) and (e) are characterized 
by  $T_{\mu\nu}\neq 0, T=const\neq 0$.

\subsection{NEC non saturating solutions}

In this case the solutions and their matter sources  must satisfy  
Eqs. (\ref{p1}), but not  Eqs. (\ref{NECsat1}) and (\ref{NEC1}).
Therefore 
they saturate the NEC of GR but not the NEC of  
nonminimally coupled theories. 
$R$ is not  constant and we get from Eq. (\ref{R})
\begin{equation}\label{R1}
R= \frac{- 6 \Box (\lambda \mathcal{L}_m f_R)+ [1 + \lambda f(R)] a}{2
\lambda \mathcal{L}_mf_R - 1}.
\end{equation}

Moreover,  the coupling function $f(R)$ is  strongly  constrained  by 
the field equations.  We get from Eq. (\ref{mattercon2})
\begin{equation}\label{fconstraint}
\frac{1}{H(r)} \frac{d^{2} H(r)}{dr^{2}}= -
\left(\frac{\lambda}
{1+ 2\lambda \mathcal L_m f_R}\right) \frac{d^{2}(\mathcal{L}_m f_R)
}{dr^{2}} ,
\end{equation}

Eqs. (\ref{fconstraint}) and (\ref{R1}) together with the other 
independent equation in (\ref{feq}) give a system of three 
equations to be solved for the unknown functions $\alpha(r)$, $H(r)$
and $f(R)$.
This system of differential equations  is very difficult 
to solve in closed form. In the following sections we will advance
some 
arguments ruling out  the existence  of  solutions of this class. 

In general the scalar curvature for this class of solutions is not 
constant. Thus, from Eq. (\ref{e1}) it follows that 
the energy-momentum tensor is not covariantly conserved.

Let us conclude this section with a remark on the uniqueness of 
black hole solutions  in NMC theories of gravity.
It is well known that in  GR Birkhoff's theorem ensures that  every
spherically symmetric 
solution generated by a charged or uncharged  pointlike mass is, up 
to spacetime diffeomorphisms equivalent to  a static solution.
Moreover, well-known no-hair theorems ensure that the SCHW and RN 
solution are unique \cite{nohair}.

By writing down the field equations of non minimally coupled gravity 
in the spherically symmetric, non-static case, one easily realizes 
that there is no evidence for a Birkhoff's theorem to hold,  i.e 
spherically symmetric solutions are not necessarily static.
For instance, one could have non static, spherically symmetric 
solutions that are not equivalent, modulo 
diffeomorphisms, to  static solutions.

\section{Cosmological constant}
\lb{SVI}

In this section we   consider a  source  described by  a cosmological 
constant, with a Lagrangian density given by Eq (\ref{b1}).
The associated  energy-momentum tensor 

\begin{equation}
T_{\mu \nu} = -\Lambda g_{\mu \nu},\quad T=-4 \Lambda,
\end{equation}
satisfies the condition (\ref{p1}), so that the considerations 
of Section \ref{sectionbh} 
hold. Following the classification of Section \ref{sectionbh}, we
distinguish 
between  
solutions saturating  or not the NEC. The first class of 
solutions gives origin to maximally symmetric spaces.

\subsection{NEC saturating solutions and maximally symmetric
spaces}

As we have seen in Section \ref{sectionbh}, analyticity of the
coupling 
function  requires $\xi=0$ in Eq. (\ref{chnulcon})  and we can 
therefore use the results of  point $(2)$ of section \ref{ff1}.

The scalar curvature is constant and determined by Eq.  (\ref{feq1})
and the field equation become the Einstein equations,
\begin{equation}\lb{ms}
R_{\mu \nu} = \tilde{\Lambda} g_{\mu \nu},
\end{equation}
with an effective cosmological constant  $\tilde\Lambda$ given by 
\beq\lb{le}
\tilde\Lambda=\frac{R_{0}}{4}=\frac{ 1+ \lambda f(R_{0}) }{1+2
\lambda\Lambda 
f_R(R_{0}) } \Lambda.
\end{equation}

We can now solve Eq. (\ref{le}) for $\Lambda$:

\begin{equation}\label{lamb}
 \Lambda =  \frac{ \tilde \Lambda}{1 + \lambda f(4\tilde \Lambda) -2
\tilde \Lambda \lambda 
 f_R( 4\tilde \Lambda)}.
\end{equation}

The solutions of Eq. (\ref{ms}) are maximally symmetric spaces 
with constant scalar curvature $R_{0}=  4 \tilde\Lambda$. Depending 
on the value of $\tilde\Lambda$ we can have the usual  three cases: 
(1) $\tilde {\Lambda}  = 0$,  Minkowski  
space-time; (2) 
$\tilde {\Lambda}  > 0$, dS space-time;
(3) $\tilde {\Lambda}   < 0$, AdS space-time.

The striking feature of the maximally symmetric spacetime  in
nonminimally 
coupled gravity  is that the curvature of the spacetime is 
determined by an ``effective'' cosmological constant  
$\tilde \Lambda$ 
which is related in a nontrivial way to the bare cosmological
constant 
$\Lambda$  through the
coupling function $f(R_{0})$. Physically this means that curvature of 
the spacetime is determined not only by the density of energy of 
matter fields, but also by the coupling function $f(R)$.
We shall get back to this important point in Section \ref{acc}.

The AdS solution takes the form
\begin{equation}\lb{ads}
e^{2\alpha}= 1+\frac{r^{2}}{L^{2}},\quad H=r,
\end{equation}
where  the AdS length $L$ is expressed in terms of the effective 
cosmological constant by  $L^{2}=-3/\tilde \Lambda$.

\subsection{NEC non saturating solutions}
\lb{sn}

In this case we have to solve Eqs. (\ref{R1}) and
(\ref{fconstraint}).
We consider only solutions  that are asymptotically  AdS, i.e  $R-R_0
\sim r^{-\gamma}$,with $\gamma>0$ and we use 
the $r\to\infty$ asymptotic expansion for 
$H(r)$:  $H(r)=\sum_{n=\alpha} a_{n}r^n\,\,$.
Under these assumptions, Eq. (\ref{fconstraint})  gives,
\begin{equation}\lb{m1}
\frac{1}{H(r)} \sum_n a_{n} n(n-1) r^{n-2} = \frac{1}{2} \left
[\frac{2 \lambda \Lambda}{1-2\lambda \Lambda f_R} \right] 
\frac{\partial^2}{\partial r^2} f_R.
\end{equation}

Assuming that $f(R)$ is analytic in $R_0$ one can write
$f_R(R)=f_R(R_0)  + f_{RR} (R_0) 
r^{-\gamma} + \mathcal O  (r^{-2\gamma}),$ so that Eq. (\ref{m1}) 
gives

\begin{equation}
\begin{split}
C\beta(\beta-1)\frac{1}{r^2} - \frac{C\beta (\beta-1)}{r^2}
\left [  2 \lambda \Lambda 
\left (f_R (R_0)+ f_{RR} (R_0) \left ( \frac{1}{r^\gamma}\right
)\right) +
\mathcal O  \left ( \frac{1}{r^{2\gamma}}\right)\right] 
& \sim \\ \sim  \lambda \Lambda  \frac{\partial^2}{\partial r^2} 
\left[ f_R (R_0)+ f_{RR} (R_0)
\left ( \frac{1}{r^\gamma}\right )+ \mathcal O \left (
\frac{1}{r^{2\gamma}}\right)\right ],
\end{split}
\end{equation}

\noindent where $Cr^\beta$ is the dominant term in $H(r)$.
One can easily realize that the only way to solve the previous 
equation is by  
setting $\beta(\beta-1)=0$,  $ (d^{2}f_R /dr^{2}) =0$,
giving back the conditions of Eq. (\ref{p10}) for NEC saturation.

This means that the only solution of the NMC theory 
asymptotically consistent with the AdS spacetime  is  the AdS 
spacetime itself.

\section {Generalized Schwarzschild solution}
\lb{SVII}
Let us now consider as  source of the gravitational field a pointlike 
particle of mass $M$. In the following, when considering pointlike 
and extended sources to conform to the usual conventions we will  use 
the Lagrangian ${\cal L}_{m}= -8\pi \rho$, where $\rho$ is the mass 
density of the source.   For the case of  a pointlike 
particle of mass $M$ the energy 
momentum tensor is therefore given by  
\beq\lb{mt}
T_{\mu \nu}=8\pi \delta_{\mu 0}\delta_{\nu 0} M\delta(r).
\eeq
In GR this gravitational source leads to the well known SCHW black 
hole solution. 

In order to derive the solutions for the  NMC  theory 
(\ref{S1}) with source (\ref{mt}), 
we first observe  that 
for $r\neq 0$, $\mathcal{L}_m=0$ and  the field equations (\ref{feq})
are the 
same as in GR. Hence the     solution has   the  same form of the
SCHW 
solution in GR,
\begin{equation}\label{c3a}
e^{2\alpha}= 1- \frac {2\tilde M}{r},\quad H=r,
\end{equation}
where $\tilde M$ is an integration constant. 

The difference between the SCHW solution of GR and that of 
the nonminimally coupled theory 
arises  only when one looks at 
the relation between 
$\tilde M$ and $M$. This   can be found, as usual, by  considering
the 
weak-field,  Newtonian limit of 
the field equations. 

The weak field limit for a  generic static, 
spherically symmetric, mass distribution with density $\rho(r)$ will 
discussed in detail in Section \ref{wf}.  Here we consider only the 
case of a pointlike source, i.e a source with a delta function 
singularity  at $r=0$. As for $r\neq 0$,  $\mathcal{L}_m=0$, we 
see that the only contributions to $R$ and $R_{00}$ come from the 
delta function singularity.
Moreover, in the usual static, nonrelativistic, weak field limit of
Eqs. 
(\ref{feq}) and (\ref{R})  do not yield the  Poisson equation unless
the 
terms containing $\mathcal{L}_m$ are identically zero. This can be 
simply achieved setting\footnote{This result can be derived along 
the same lines used in  Section \ref{wf} to derive the conditions for 
the existence of the Newtonian limit.}
\beq\lb{pl}
f_{R}(R(r=0)=0.
\eeq

Because $R\propto \delta(r)$, the previous Eq. (\ref{pl}) is 
equivalent to $f_{R}(R \to\infty)=0$.

Using  Eq. (\ref{pl}) we can perform the usual static, 
nonrelativistic, weak field limit in the field equations (\ref{feq})  
and (\ref{R}) and obtain Poisson equation for the  
Newtonian potential $\Phi$,
\begin{equation}\lb{g4}
\nabla^2 \Phi = \frac{1}{2}\left[1+ \lambda f(R\to\infty)\right]
T_{00}.
\end{equation}

\noindent Whereas for GR,  $f(R)=0$, and  Eqs. (\ref{c3a}), (\ref{g4})
give $M=\tilde 
M$; for the NMC theory, one
finds
\begin{equation}\label{mass}
\tilde M=\left[1+\lambda f(R\to\infty))\right]M.
\end{equation}

$\tilde M$ has to be considered as the ``effective'' dressed
gravitational mass of the black hole, 
whereas $M$ is the "bare" one. In order to the two masses be the same:
\begin{equation}\label{c4}
f(R \to \infty)=0.
\end{equation}
Notice that the relation
(\ref{mass}) yields a  
divergent mass if $f(R)$ diverges for $R\to \infty$.

Naively, one is lead to  interpret $\tilde M$ as the gravitational
mass 
and $M$ as the inertial mass of the pointlike particle and the
relation 
(\ref{mass}) as an explicit breakdown of the equivalence principle.
However, closer inspection reveals that this  cannot be simply 
inferred by considering the motion of a test particle in a SCHW 
background.
In fact, although it is true that equation of motion of a test 
particle of mass $m$ in the SCHW metric contains a $f(R)$-dependent
connection,  the 
$R$-dependent part cancels because $R=0$, identically, whenever 
$r\neq0$ and the motion is purely geodesic:

\begin{equation}\lb{h4}
\frac{\delta S_m}{\delta x^{\mu}(\tau)}= m(1 + \lambda f (0) ) 
 \frac{\delta}{\delta x^{\mu}(\tau)} \int ds=0.
\end{equation}
We see that the coupling function does not effect
the geodesic motion of the test particle. 
Moreover, Eq. (\ref{h4}) seems to tell us that the ``inertial'' mass 
of the test particle  is equal to the dressed mass $\tilde m=m(1 +
\lambda f (0) ) 
$.  As long as we consider the motion of a test particle in a SCHW 
background there is no physical way to distinguish between $M$ and 
$\tilde M$. 

The distinction between bare and dressed mass and  the observation of 
breakdown of the equivalence principle requires to go beyond the test 
particle approximation (for instance by considering the mutual 
interaction of two black holes),  or to consider more general matter 
distributions (for instance a collapsing star). In this general 
situations $R\neq const$ and the equation of motion for matter 
(\ref{h4}) will get a $f(R)$-dependent contribution to the connection.

\subsection{Generalized SAdS solution}

The  asymptotically flat generalized SCHW solution (\ref{c3a}) can be 
easily extended to the case  in which a negative cosmological constant
$\Lambda$ is present.
In GR one  obtains in this way the asymptotically AdS black hole
solutions 
(SAdS solutions).

Considering the case where the  NEC is saturated, the SAdS solution
is easily
obtained  
from the AdS solution (\ref{ads})  by introducing a matter 
source of the form (\ref{mt}):
\begin{equation}\label{c3}
e^{2\alpha}= 1+ \frac{r^{2}}{L^{2}}- \frac {2\tilde M}{r},\quad H=r,
\end{equation}
where $\tilde M$ is the effective gravitational mass (\ref{mass}) and 
$L^{2}=-3/\tilde \Lambda$, $\tilde \Lambda$ being  the effective 
cosmological constant  (\ref{le}).

\section{Charged black hole  solutions}  
\lb{SVIII}
In this section, we consider electromagnetically charged black hole
solutions in the 
 NMC theory.
The matter-field  action  is obtained by using in Eq. (\ref{S1})  the 
Lagrangian density for the electromagnetic (EM)
field:
 \begin{equation}\label{lagem}
 \mathcal{L}_m = - F^2 =- F_{\mu \nu}F^{\mu \nu}.
 \end{equation}
The $F^{\mu \nu}$ is the EM tensor.
The resulting  field equations for the EM field  are
\begin{equation}\label{maxwell}
\frac{1}{\sqrt{- g}}\partial _\mu(\sqrt{-
g} F^{\mu \nu} (1+ \lambda f(R)))=0.
\end{equation}
We first  consider the  purely electrically charged case. Only
the $tr$  component of the
electromagnetic tensor is non vanishing  and the Maxwell equations 
(\ref{maxwell})
can be immediately
solved to give
 \begin{equation}\label{f1}
 F_{01} = -F_{10} = -\frac{Q}{H^2 (1+\lambda f)},
 \end{equation}
where $Q$ is an integration constant, which in GR gives the electric 
charge of the  solution.
The energy-momentum for the EM field is given by,
\begin{equation}\lb{emt}
T_{00} = 4 \frac{Q^2}{H^4 (1+\lambda f)^2} e^{2 \alpha},\quad
T_{11} = -  e^{-4 \alpha}T_{00},\quad
T_{22} = 4 \frac{Q^2}{H^2 (1+\lambda f)^2} ,\quad
T_{33}  = \sin^2 \theta T_{22}.
\end{equation}

Moreover using Eq. (\ref{f1}) one easily finds that
\begin{equation}\label{h2}
 \mathcal{L}_m = - F^2=2 (F_{01})^2 =  \frac{2Q^2}{H^4 (1+\lambda
f)^{2}}.
 \end{equation}
It can be checked that the energy momentum tensor (\ref{emt}) 
satisfies the conditions   $T=0$ and $ T_{0}^{0}= 
T_{1}^{1}=0$, which is a particular case of those discussed in
Section 
\ref{sectionbh}. The first condition expresses the conformal 
invariance of the EM action, while the second concerns the GR NEC
saturation.

As before, we distinguish between the solutions that saturate NEC
from the ones that do not in the NMC theory. 

\subsection{NEC  saturating solutions}
\lb{ssn}

The  general results of Section \ref{ff1} tell us that solutions 
characterized by $\xi\neq 0$ are ruled out due to the mismatch of the
asymptotics 
and of the analyticity of 
$f(R)$. For  charged solutions the conditions ruling out $\xi\neq 0$  
hold even more strongly because from Eq. (\ref{h2}) we see that as
$r\to 
\infty $,  $\mathcal{L}_m$ decays  as $\mathcal{O}(r^{-4})$.

NEC saturating solutions are therefore  obtained setting 
$\xi=0$ and using Eqs. (\ref{h2}) and (\ref{k2}). 
For $b=T=0$, we get from  Eqs. (\ref{h2}) and (\ref{k2})
\begin{equation}\label{con}
R=0,\quad f_R(0)=0.
\end{equation}

From this, the field equations (\ref{feq}) give
\begin{equation}\label{eql}
 R_{00}= -
\frac{4 g_{00 }Q^2}{r^4 (1+ \lambda f(0))},\quad
\frac{d}{dr}\left [ r e^{2 \alpha} \right ] = 1- \frac{4 Q^2}{r^2(1+
\lambda f(0)} .
\end{equation} 
If we  take  $\lambda = 0$   we
recover, as expected, 
the GR equations,
\begin{equation}\label{eqm}
 R_{00}= -
\frac{4 g_{00 }Q^2}{r^4},\quad
\frac{d}{dr}\left [ r e^{2 \alpha} \right ] = 1- \frac{4 Q^2}{r^2} .
\end{equation} 

The system of Eqs. (\ref{eql}) becomes completely equivalent to
(\ref{eqm}) 
by defining an ``effective'', dressed, charge 
\begin{equation}\label{kk}
\tilde Q=\frac{Q}{
\sqrt{1+\lambda f(0)}}.
\end{equation}
It follows immediately that the  charged black solutions of Eqs. 
(\ref{eql}) have the same form of the RN solution of GR, with the 
charge $Q$ and mass $M$ replaced respectively by the effective charge
$\tilde Q$ 
of Eq. (\ref{kk})  and by the effective mass $\tilde M$ of Eq. 
(\ref{mass}):
\begin{equation}\label{metric}
ds^2 =  -\left (1-\frac{2\tilde M}{r} + \frac{4\tilde Q^2}{r^2
}\right 
)dt^2 + \left (1-\frac{2\tilde M}{r} + \frac{4\tilde Q^2}{r^2
}\right )^{-1}dr^2 +r^2 d \Omega^2 .
\end{equation}

The dressed and bare charges are 
the same when

\beq\lb{pp1}
f(R=0)=0.
\eeq

The electrically charged solution of the
NMC gravity theory can be simply obtained from the 
RN black hole solutions of GR by a  simple, coupling function 
dependent, charge and mass  rescaling.
This means that the physical effect of the nonminimal coupling on the 
charged static black hole solution is a  redefinition of the 
physical parameters of the black hole, which  
contains the information about the local  behaviour of the  coupling
function
at $R=0$ and $R=\infty$. 

On the other hand, we see that  solution (\ref{metric}) is allowed
only  if the
coupling function $f(R)$ is 
constrained by Eq. (\ref{con}). This is an important selection  
criteria for the coupling function. 

It should be pointed out that, similarly to the dressed mass, as long
as one considers only test 
particles  the physically observable charge of the black hole is
the dressed one $\tilde Q$ of Eq. (\ref{kk}). This is immediately 
evident from the solution for the Maxwell field strength (\ref{f1}).

It is also of interest to compare the behaviour of charged black hole 
solutions of the NMC gravity theory with the 
charged black hole solutions of the 
Dilaton-Maxwell-Einstein gravity. In the latter theories, the 
charge rescaling effect is very similar to Eq. (\ref{kk}), with the
rescaling 
factor  determined by the asymptotic value of the dilaton 
\cite{dila,dila1}.
However,
both 
the form and the causal structure of these solution are radically 
different from the RN solutions. Strangely enough the NMC 
 theory allows for charged solutions which are a ``minimal'' 
deformation (just a charge and mass rescaling) of the RN solutions.  

The occurrence of this  minimal deformation  can  be also directly 
seen from Eq. (\ref{e1}). The stress energy tensor is covariantly 
conserved  whenever  $R=0$. 

\subsection {Dual and dyonic solutions}
 Electromagnetic duality holds also in the 
NMC theory (\ref{S1})  with the matter Lagrangian 
density given by (\ref{lagem}). In fact, the transformation $F\to
\tilde
F$ 
with $\tilde F^{\mu \nu} =\epsilon ^{\mu \nu \rho \sigma} F_{\rho 
\sigma}$   leaves  the action (\ref{lagem}) unchanged 
up to  a sign and leads,  therefore, to the same Maxwell
equations. 
The duality transformation leaves the metric part of the solution 
unchanged and transforms the purely electric solution 
(\ref{f1}) into a purely magnetic monopole solution

\begin{equation}
F_{23}= \frac{P}{1 + \lambda f}\sin \theta,
\end{equation}
where $P$ is the magnetic charge. We can define, analogously to Eq. 
(\ref{kk}), the effective, dressed, magnetic charge
\begin{equation}\label{km}
\tilde P=\frac{P}{
\sqrt{1+\lambda f(0)}}.
\end{equation}
For sources carrying both electric and magnetic charges we have the 
dyonic
charged  black hole solution given by,
\begin{equation}\label{generalmetric}
ds^2 =  -\left (1-\frac{2\tilde M}{r} + \frac{4(\tilde Q^2+\tilde
P^2)}{(r^2 }\right )dt^2 + 
\left (1-\frac{2 \tilde M}{r} + \frac{4(\tilde Q^2+\tilde P^2)}{r^2}
\right )^{-1}dr^2 +r^2 d \Omega^2.
\end{equation}

\subsection{Generalized RNAdS solutions }

Similarly to the SAdS solution one can easily construct 
the generalization of the RNAdS 
solution of GR for  the  NMC theory. 
The matter Lagrangian density has 
now (apart  from the mass $M$ of the source localized in $r=0$) two
contributions, 
a cosmological constant and an electromagnetic 
term:
\begin{equation}\label{lagrangian}
 \mathcal{L}_m = -\Lambda- F^2.
 \end{equation}
The corresponding stress-energy tensor $T_{\mu\nu}$ will have two 
contributions and the purely electric,  static, spherically symmetric
solution, 
with $\xi=0$, is given by 
\begin{equation}\label{c3c}
e^{2\alpha}= 1+ \frac{r^{2}}{L^{2}}- \frac {2\tilde M}{r}+\frac{4
\tilde Q^2}{r^2 },\quad H=r,
\end{equation}
where $\tilde M$ is the effective gravitational mass Eq.
(\ref{mass}), 
$L^{2}=-3/\tilde \Lambda$, $\tilde \Lambda$ being  the effective 
cosmological constant Eq. (\ref{le}), and $\tilde Q$ is the
effective 
charge Eq. (\ref{kk}).

\subsection{NEC non saturating  solutions}

The existence  of  charged solutions that do not saturate NEC can be 
investigated using a method similar to that used in Section \ref{sn}.
We search for asymptotically  flat solutions of  Eqs. (\ref{R1}) and
(\ref{fconstraint}) 
with   $\mathcal{L}_m$ given by Eq. (\ref{lagem}).
We consider therefore  $R
\sim r^{-\gamma}$,with $\gamma>0$ and we use 
the same $r\to\infty$ asymptotic expansion for 
$H(r)$, namely,  $H(r)=\sum_{n=\beta} a_{n}r^n,\,\,$.

Assuming that $f(R)$ is analytic in $R=0$, Eq. (\ref{fconstraint})
gives at leading order in $r\to \infty$:

\begin{equation}\label{alpha1}
\begin{split}
\frac{\beta(\beta -1)}{r^{2 }} +\frac{4 \lambda Q^2 \beta
(\beta-1)}{r^{4 \beta + 2}} \left [ \frac{f_R(0)}{[1+\lambda
f(0)]^2} \right ]  + \mathcal {O}\left (\frac{1}{r^{\gamma+4 \beta +
2}}
\right)& \sim \\
 \sim - 2    \left [ \frac{20\lambda Q^2}{r^{4 \beta + 2}}
\frac{f_R(0)}{[1+\lambda f(0)]^2}  + \mathcal {O}\left
(\frac{1}{r^{\gamma
+4 \beta + 2}} \right) \right].
\end{split}
\end{equation}
It follows immediately that, at leading order, we must have  $\beta
=1$, $f_R(0)=0$,
which leads to the same solutions obtained  in Section \ref{ssn} when
NEC was saturated.

\section{Weak field limit}
\lb{wf}
In this section we discuss the weak field limit of the NMC
 theory  (\ref{S1})  for the case of a static, 
spherically symmetric, extended source with rest mass density
$\rho(r)$.
Differently from GR, the weak-field limit of the  gravity  theory
(\ref{S1}) is 
a rather  involved issue. This is due to the fact that 
its feature 
depends crucially on the form of the coupling function $f(R)$.  This
is 
immediately evident if one considers  the weak-field expansion near 
flat space. If the coupling function $f(R)$ diverges in $R=0$ this 
weak field limit is meaningless. 
It is also possible that non minimally coupled gravity theories allow 
for alternative weak coupling expansions, for instance, near some  
background with nonvanishing curvature.

In this paper we assume  the validity of the usual weak-field 
expansion around a flat space.  We circumvent the above problems by 
assuming that the coupling function is 
analytic in
$R=0$, so that we can write,
\begin{equation}
f(R) = f (0) + f_R(0) R + \mathcal{O}(R^2).
\end{equation}

We then use  the usual weak-field, nonrelativistic, expansion of  the
metric around flat 
space-time
\beq\lb{gg1}
g_{\mu \nu} \approx \eta_{\mu \nu} + h_{\mu \nu},
\end{equation}
where $\eta_{\mu \nu}$ is the  flat space-time metric tensor and
$\left | h_{\mu \nu}\right| <<1$. We   assume as usual   that $h_{\mu
\nu}$ is 
time independent, that the matter Lagrangian density is dominated by
the 
mass density, $\mathcal{L}_m  =- 8\pi  \rho (r)$ and that the
density 
vanishes asymptotically, $\lim_{R\rightarrow 0}
\rho =0$.

In the weak field, nonrelativistic  limit, the field equations are
dominated
by the the rest mass, so that the only relevant part is the
$tt$-component.
Expanding the coupling function $f(R)$ in power series, retaining 
only the linear terms in the curvature and using the following field
expansions,
\begin{equation}
 \nabla_\nu \nabla_\mu [  \rho (f_{RR} (0) R )] \approx
\partial_\mu \partial_\nu  [ \rho (f_{RR} (0) R )],
\quad g_{\mu \nu} \Box [ \rho (f_{RR} (0) R )] \approx
\eta_{\mu
\nu} (f_{RR} (0)\frac{d^2(\rho  R )}{dr^2} ,
\end{equation}

we get from the field equations (\ref{feq}).

\begin{equation}\lb{w1}
[1 - 16 \pi\lambda  \rho f_R(0)] R_{00} -
\frac{1}{2}g_{00} R  = -2 \lambda \frac{d^2}{dr^2} \left[ 8
\pi\rho
(f_R(0)+f_{RR}
(0) R )\right]+  8 \pi[1 +
\lambda(f (0) + f_R(0) R) ] \rho (r).
\end{equation}

We now compute from Eq. (\ref{R}) the scalar curvature in our 
approximation. We get  
\begin{equation}
R \approx \frac{1}{1+ 16 \pi\lambda \rho f_R (0)} \left 
\{ 8 \pi (1+\lambda
f(0)) \rho - 48 \pi
\lambda f_{RR}(0) \frac{d^2 (\rho R)}{dr^2} - 48 \pi\lambda
f_{R}(0)
\frac{d^2\rho} {dr^2}\right\}.
\end{equation}
 Retaining only the linear terms in  $\rho$ we have 
\begin{equation}
R \approx 8 \pi\left(1+\lambda f(0)\right)\rho - 48
\pi\lambda f_{R}(0)
\frac{d^2\rho} {dr^2}.
\end{equation}
Using the usual static, non relativistic, weak field 
expression  $R_{00} \approx -\frac{1}{2} \nabla^2 h_{00}$, to leading 
order in the curvature  and in  $\rho$,
Eq. (\ref{w1}) 
becomes
\begin{equation}\lb{ll1}
 \nabla^2 h_{00}\approx 
 - 16 \pi \lambda  f_R(0)\frac{d^2\rho} {dr^2}-8
\pi\left[1 + \lambda 
 f (0)\right] \rho (r).
\end{equation}
We see  that for  generic coupling function $f(R)$ and generic matter 
distribution   the previous equation is not the usual
Poisson equation, i.e  the weak 
field limit of the NMC theory is not the usual
Newtonian 
gravity.  If we exclude the  particular matter 
distribution $\rho(r)= a r +b$ (not compatible with the boundary 
condition $\lim_{R\rightarrow 0}
\rho =0$),
 the only way to recover the
 Newtonian limit is to constrain the value of $f_{R}(0)$ 
as    in  Eq. (\ref{con}), i.e  $f_R(0)=0$. 

Once the constrain Eq. (\ref{con}) is used, Eq. (\ref{ll1}) takes the 
usual Poisson form 
\beq\lb{po}
\nabla^2 h_{00}  = - 8 \pi \tilde \rho (r),
\eeq
if one 
defines a coupling function dressed mass density given by 
\begin{equation}\lb{hh1}
\tilde \rho =  [1 + \lambda f (0) ] \rho (r) .
\end{equation}

The condition for the equality  of the dressed and bare  rest 
mass densities is $f(0)=0$, i.e the same conditions (\ref{pp1}) for 
the equality  of dressed and bare black hole charges.
This is rather unexpected.  Completely uncorrelated results, 
namely the existence of charged black hole solutions and the equality 
$Q=\tilde Q$ on one side, and the 
existence of the usual Newtonian limit and the equality $\tilde 
\rho=\rho$ on the other side, can be achieved using the same 
conditions (\ref{con}), (\ref{pp1}).

Notice that both the conditions for the existence of the usual weak 
field Newtonian limit  Eq. (\ref{con}) and the definition of dressed
mass 
density (\ref{hh1}) are, respectively, 
analogous to  the condition for the existence of the Newtonian limit 
for a pointlike source (\ref{pl})  and to the definition of dressed
black hole mass 
given by Eq. (\ref{mass}). There is, however, a crucial and important 
difference.  For the pointlike particle, the existence  conditions of 
the Newtonian limit  constrain the 
coupling function in the UV (at $r=0$ or equivalently at $R\to\infty$)
and the dressing of the black 
hole mass is realized in terms of the behaviour of the coupling 
function $f(R)$ in the same UV region. 
On the other hand for a spherically symmetric extended  source 
considered in this section, the condition of existence and the 
dressing are characterized by the behaviour of the coupling function 
in the IR (i.e at $r\to\infty$ or equivalently, $R=0$).
This is a rather natural feature if one considers the localization of 
the pointlike source.

\section{ The coupling function}
\lb{cf}

The NMC gravity theory (\ref{S1}) has to be
considered as an 
effective description of some, yet to be discovered, fundamental 
theory of gravity.  The coupling function contains  
presumably information about  the behaviour of the fundamental theory 
in its various dynamical regimes. In the previous sections we have 
found that the existence of black holes solutions and the form of the
physical 
parameters associated to them gives local constraints on the coupling 
function $f(R)$. This  information can be used both  for selecting a 
form of the coupling function $f(R)$ or, eventually, falsify the
theory.

It is also important to stress that the information about the
coupling 
function $f(R)$ inferred from the existence of black holes
has a strong heuristic power as it covers the full range of 
energy scales  of the gravitational interaction.  In fact, we have 
both constraints on $f(R)$ in the, $R\to\infty$, strong curvature
region (corresponding 
to the ultraviolet, quantum gravity regime) and constraints on
$f(R)$ for, 
$R=0$,  weak curvature region (corresponding 
to the infrared regime)   of the gravitational interaction.

The information about $f(R)$ that can be gathered from our previous 
results about static, spherically symmetric solutions  of the
NMC theory can be 
organized in two classes:  $(a)$ Constraints on $f(R)$ emerging from
the requirement of existence of solutions; $(b)$ constraints arising
from 
the form of black hole dressed parameters.

\subsection{Constraints on \texorpdfstring{$f(R)$}{f(R)} arising from the 
existence of solutions}
\lb{sdressaa}

Existence of the Minkowski space vacuum solution  does not 
set any constraint on $f(R)$. On the other hand, from Eq. (\ref{le}) 
the existence of the   dS and AdS solution constrains the IR 
behaviour of the coupling function $f(R)$, which must 
be finite, as well as its first derivative, when evaluated on $R_{0}$
(spacetime curvature corresponding
to 
the  cosmological constant).

The existence of the SCHW solutions constrains the UV behaviour of 
$f(R)$. From Eq. (\ref{pl}) and  (\ref{mass}) we see that 
$f_{R}(R\to\infty)=0$ and  $f(R\to\infty)$ must be finite.
On the other hand, the existence of the RN  solutions (\ref{metric})
requires
that $f_{R}(R=0)=0$, and from (\ref{kk}), that $f(R=0)$ is finite,
constraining the deep IR behaviour of the coupling function.

Considering both pointlike and extended, spherically symmetric 
sources, the existence of the Newtonian weak field limit constrains
the 
coupling function $f(R)$  both at the UV and at IR. For pointlike
sources,
we 
have $f_{R}(R\to\infty)=0$. For extended sources we have
$f_{R}(R=0)=0$, 
the same conditions for the existence of the RN solution.

\subsection{Constraints arising from 
the  dressed parameters}
\lb{sdress}
The existence of the relations  (\ref{mass}), (\ref{kk}),
(\ref{le}) and (\ref{hh1})
between bare and dressed black hole observables has a strong
predictive power.
In fact, by measuring  gravitationally dressed parameters $\tilde
\Lambda,
\tilde M, \tilde 
 Q,\tilde \rho$ and the bare ones one could either get informations
on the coupling function $f(R)$ or  falsify the theory.

Assuming that we can measure independently both the dressed and bare 
parameters, these relations can  be used to restrict the  form of the 
coupling function $f(R)$ at different dynamical regimes.

Let us consider as an example the situations in which  
all the 
standard spherically symmetric solutions of GR (SCHW, RN, dS, AdS,
SADS, 
RNAdS) fully coincide with those of the NMC theory, 
i.e  all the dressed parameters turn out to be  
equal to the  bare ones.  A solution of the  equations  $M=\tilde 
M,\, Q=\tilde Q,\, \Lambda= \tilde \Lambda, \tilde \rho=\rho$ is given
by 
\bea\lb{kkk1}
f_{R}(R\to\infty)&=&0,\quad f(R\to\infty)=0,\quad
f_{R}(R=0)=0,\nonumber\\
f(R=0)&=&0,\quad 
f_{R}(R=R_{0})=0,\quad  f(R=R_{0})=0.
\eea
Eq. (\ref{kkk1}) give conditions on the local behaviour of the 
coupling function $f(R)$ at  different length scales.  The first 
two conditions, which arise from existence of the SCHW solution and
from 
$M=\tilde M$, essentially tells us that  
quantum gravity effects can be described by   the NMC theory  as a
perturbation of GR. 
Conversely, the  third and fourth  conditions in (\ref{kkk1}),
arising 
from the existence of the RN solutions and from $\tilde Q=Q$,  imply 
 the same, but in the IR: any infrared modification of GR can be 
formulated as a perturbative expansion of GR.

Finally, the fifth and sixth  conditions in (\ref{kkk1}),  arising
from $\tilde 
\Lambda= \Lambda$ allows us  to expand perturbatively the  
NMC theory  near GR  at some, intermediate,  
cosmological length scale.

An  observation  of a mismatch  between  bare and dressed 
parameter would be a strong indication of the validity of the
nonminimally 
coupled theory.

Notice that the solution of the equation $\Lambda=\tilde\Lambda$
given 
by the last two equations in (\ref{kkk1})  is not unique. In fact, 
using (\ref{lamb}) one finds that  
this equation is generally solved by $2f(R_{0})= R_{0}f_{R}(R_{0})$.

Apart from the existence and behaviour of black hole solutions,
the 
form of the coupling function $f(R)$ can be constrained by several
phenomenological  considerations. Two of them are of
particular 
relevance 
 $(a)$ Experimental constraints on 
the post-Newtonian parameters; $(b)$ In order to keep gravity
attractive, assuming $\lambda
f(R) <<1$, then $
 f_R(R) < \frac{1}{2 \lambda}$. 

The experimental constraints on the post-Newtonian parameter have
been 
discussed in Ref. \cite{OBMarchParamos} starting with a more
general action
given by: $S= \int [f_1(R) + f(R) \mathcal L ] \sqrt{-g}d^4 x$ and 
assuming  definite, albeit rather general form for the coupling 
function $f(R)$.

As an example of implementation of the previous conditions,
we consider here a generalization  of the coupling function 
proposed in Ref.  \cite{OBMarchParamos}, which is suitable to 
satisfy the conditions for the existence of the various  black hole
solutions
of the nonminimally coupled theory. We 
take
\begin{equation}\label{fR}
f(R) = \sum _{n>0}^{N} \frac{r_n^{2n}}{R^{2n} +C_n} +K,
\end{equation}
where  $r_{n},\,C_n$ and $K$  are parameters, and $K= f (R \rightarrow
\infty)$.\\

One can easily see that this form of $f(R)$ identically satisfies the 
condition for the existence of the SCHW solution (\ref{pl}),
the RN
solution (\ref{con}) (hence also for the existence of the Newtonian 
limit for an extended, spherically symmetric source). 
Moreover,  using the coupling function (\ref{fR})  the conditions 
(\ref{kkk1}) for  the equality between  dressed and bare parameters
can be easily 
implemented in terms of the relation between the coefficients
appearing 
in Eq. (\ref{fR}).

The equality between $\tilde M$ and $M$ can be easily obtained 
setting $K=0$.
On the other hand, the equality  between $\tilde Q $ and $Q$ (hence 
the equality between $\rho=\tilde \rho$)  requires
\beq\lb{pp1a}
f (0) = \sum_n \frac{r_n^{2n}}{C_n} + K=0.
\end{equation}
Finally, introducing  Eq. (\ref{fR}) into Eq. (\ref{le}) one finds
that
the equality
$\tilde \Lambda $ and $\Lambda$ can be satisfied by requiring
\begin{equation}
  C_n =-(n+1) R_{0}^{2n}.
\end{equation}

\vspace{1cm}
\subsection{Other possible scenarios}
\lb{ops}
The scenario  described in Section \ref{sdress} is a rather 
conservative one. $f(R)$ is assumed to be analytic  and  all 
the conditions for the existence of static 
spherically symmetric solutions and of the Newtonian limit  are
satisfied and, additionally,  the 
conditions (\ref{kkk1}) for the equality of bare and dressed 
parameters also hold. Thus, at all the relevant physical scales (UV, 
cosmological, far IR) the gravitational interaction 
can be described  as perturbative  expansion of 
the nonminimally coupled theory (\ref{S1}) in terms of powers of 
the curvature,  with   GR  being 
the leading approximation.

Obviously, this is not the only logical possible scenario. There are 
several other possibilities.
The first  one is that we can retain analyticity of  
$f(R)$, but  at least some of the equalities (\ref{kkk1}) are 
not satisfied. If this is the case, we still have a perturbative 
expansion  around GR, but we would observe a modification of the 
spherically symmetric solutions of GR and/or of the weak field limit.

The other possibilities are related to a failure of analyticity of 
$f(R)$ at some dynamical regime. This failure can occur $(a)$ at the 
UV (at $r=0$); $(b)$ at some intermediate scale $l$; $(c)$ in the far
IR
($r\to\infty$). 
One of the most important results of this paper is the claim that 
this failure does have a direct impact on the 
existence of static, spherically symmetric solutions.
If $f(R)$ diverges at $R\to\infty$ this implies that 
the SCHW solution in the nonminimally coupled theory does nor exist.
Physically,
this means that  
if quantum gravity effects cannot be described, perturbatively,  by a 
theory (\ref{S1}) the SCHW black hole solution cannot exist.
Conversely,  a divergence of $f(R)$ in the far IR (i.e at
$r\to\infty$), implying that we 
cannot describe, perturbatively, infrared modification of gravity  
using a theory (\ref{S1}), would imply that both  the RN solution and 
of the usual Newtonian limit are not admitted.

The most interesting application of  the gravity theory (\ref{S1}) is 
to describe infrared modifications of gravity in order to mimic
dark matter \cite{DM mim1, DM mim2}.  This is the case described by
point  $(b)$ above. 
We generically require 
analyticity of  $f(R)$ at some intermediate scale. An interesting 
choice for $f(R)$ is an inverse power law \cite{DE mim1}:
\beq\lb{ip}
f(R)=\left(\frac{R_{s}}{R}\right)^{n},\quad n>0.
\eeq
The coupling function (\ref{ip}) is analytic at every intermediate 
scale, for which the curvature $R=R_{0}$, but it diverges for $R=0$.
Therefore, one can naively expect that both the RN and the Newtonian
limit, 
for an extended spherically symmetric source, not to exist.
However, this may not be necessarily the case.   First, the  
form (\ref{ip}) could be  an approximation that holds near $R=R_{0}$, 
whereas the exact form of $f(R)$ could be perfectly regular at $R=0$.
Second,  all the results of this paper about the weak field limit 
have been derived  assuming the usual weak field expansion around 
flat space ($R=0$). An alternative weak field expansion, near some 
curved background can yield weak field features distinct from
the ones discussed in this paper (see eg. Ref.
\cite{OBMarchParamos}).

\section{Application  to the cosmological constant problem}
\lb{acc}
In this section we consider the dressing of the cosmological constant 
through the coupling function $f(R)$ given by Eq. (\ref{le}) to 
address the cosmological constant problem \cite{cos,cos1,cos2,cos3}.

In its standard formulation, the cosmological constant problem is the 
difficulty to explain why the present value of the cosmological 
constant, inferred from the  universe acceleration data, is 120 
orders of magnitude smaller than its natural value, inferred from
microscopic 
physics ($\Lambda\sim M_{p}^{2}$, $M_{p}$ being the Planck mass).

The dressing of the cosmological constant (\ref{le}) in the 
NMC does seem to provide a natural way to adjust 
the cosmological constant at the level of the effective theory of 
gravity. The nonminimally coupled theory has to be considered as
an effective theory and   the coupling function $f(R)$ 
encodes the information about the gravitational  dressing of all the 
physical parameters. In particular, in Eq. (\ref{le}), the bare 
cosmological constant $\Lambda$ describes the vacuum energy of matter 
fields, whereas the dressed cosmological constant $\tilde \Lambda$ is 
the effective gravitating vacuum energy of the matter fields.

The actual value of $\Lambda$ depends on microphysics, it can range
from the 
(TeV scale)$^{2}$ to $M_{p}^{2}$. Here we assume $\Lambda\sim 
M_{p}^{2}$.

The key question, which we address in this section is: is there a 
choice of the coupling function $f(R)$, which on the one side allows
to 
achieve in a natural way $\Lambda = 10^{120}\tilde \Lambda$  and on 
the other side is compatible with the constraints discussed in
Section \ref{cf}.

By natural, we mean that the hierarchy of scales is achieved   
without any fine tuning and without introducing any other mass scale 
apart from $\tilde \Lambda$ (or $\Lambda$).
It should also be noted that the constraints discussed in Section 
\ref{cf} are only a subset of the full set  
of conditions that the coupling function must satisfy such as, for
instance, 
consistency with solar system data,  the post-Newtonian  
approximation, gravitational lensing  and so on. For sure, this is 
a limitation of the discussion 
of this section about  the form of $f(R)$.  

We first point out that a coupling function of the form (\ref{ip})
does 
not do allow for generating a ratio (see also discussions in Refs.
\cite{Lambda mim,modfriedmann}),
\beq\lb{gf}
\frac{\Lambda}{\tilde \Lambda}\sim 10^{120}.
\eeq
In fact, introducing Eq. (\ref{ip}) into Eq. (\ref{lamb}), one obtains
\beq\lb{ff}
\frac{\Lambda}{\tilde \Lambda}=\frac{1}{1+\lambda 
(1+\frac{n}{2})\left(\frac{R_{s}}{4\tilde\Lambda}\right)^{n}},
\eeq
which for a positive $\lambda$ is evidently incompatible with 
(\ref{gf}).
 The situation  is not much better if we take instead of  Eq.
(\ref{ip}) 
 $f(R)$ as a positive power $f=(R/R_{s})^{n}$. 
In this case, we get the same Eq. (\ref{ff}), but with the opposite
sign of 
$n$.  Thus, for $n>2$ (\ref{ff}) becomes compatible with Eq. 
(\ref{gf}) but, nevertheless accurate fine tuning is needed to 
achieve  (\ref{gf}). For instance, for $n=4$ we obtain  ($\lambda$ is 
here irrelevant and has been set to $1$),
$1-(4\tilde \Lambda/R_{s})^{4}= 10^{-120}$. Accurate fine 
tuning of the parameter $R_{s}$ is needed in order to satisfy this 
equation.

Similar arguments rule out coupling functions  with polynomial or 
rational form like (\ref{fR}), leaving as possible candidates 
coupling functions with exponential form.

Let us  assume that in the region of small
spacetime curvatures $ 
R\sim \tilde\Lambda$   
the  coupling function $f$ is well described by the function:
\begin{equation}\label{h1}
\lambda f(R)=e^{-\sigma R}-1,
\end{equation}
where $\sigma$ is a positive parameter to be determined in order to
achieve 
the hierarchy (\ref{gf}).
This coupling function satisfy Eqs. (\ref{pl}),
 but not Eq.  (\ref{c4}) and (\ref{con}). 
As such, it allows for the existence of the SCHW solution but not for 
the existence of the  RN solution and of the Newtonian limit for an 
extended, spherically symmetric source. Moreover,  because Eq. 
(\ref{c4}) is not satisfied, the dressed mass of the SCHW  black hole 
does not coincide with its bare mass.

One can easily show that the coupling function (\ref{h1}) allows 
for generating  the hierarchy (\ref{gf}) without any unnatural fine
tuning of 
the parameter $\sigma$.
In fact, introducing 
Eq. (\ref{h1}) into Eq. (\ref{lamb}) one gets,
\begin{equation}\label{lamb1}
 \frac{\Lambda}{ \tilde \Lambda} =  \frac{ e^{4\sigma\tilde\Lambda}}{1
+2\sigma\tilde\Lambda}.
\end{equation}
Choosing in Eq. (\ref{h1}) the dimensional constant $\sigma$ as
$\sigma\approx 
70/\tilde\Lambda$, i.e  just two orders of 
magnitude bigger then $1/\tilde\Lambda$,  one managers to satisfy Eq.
(\ref{gf}).

The physical mechanism behind the generation of the hierarchy 
(\ref{gf}) can be also understood by calculating the effective 
gravitational coupling $\kappa$, Eq. (\ref{ecc}).
In the region of small spacetime curvature, where the form of $f(R)$
(\ref{h1}) holds,  we naturally 
expect the density of usual matter to be one or two orders of 
magnitude smaller then that of dark energy. We can therefore  
realistically assume  $\mathcal{L}_{m}\sim 10^{-2}\tilde \Lambda$ so 
that we have,

\begin{equation}\label{ecc1}
\kappa\sim (1+\lambda f(R_{0}))\sim e^{-4\sigma \tilde\Lambda}\sim
10^{-120}.
\end{equation}
This means that the effective cosmological constant is so  tiny 
compared  to the expected vacuum energy of matter fields, 
because the effective coupling constant (\ref{ecc}) becomes extremely 
tiny in the regions of small spacetime curvature.

Although appealing from several points of view, the choice of the
coupling function 
(\ref{h1}) is not completely 
satisfactory. The main problem is that it does not allow for the 
existence of the RN solution and of the Newtonian limit for 
extended, spherically symmetric sources. This is particularly 
disturbing because the coupling function (\ref{h1}) is assumed to be
a good description in the regions of small $R$, the same region 
relevant for the above existence conditions.

We can do  better by assuming that in the region of small
spacetime curvatures, $f$ is given by

\begin{equation}\lb{op1}
\lambda f(R) =e^{ -\sigma^2 R^2}-1.
\end{equation}
This choice for the coupling function allows  to satisfy not only Eq.
(\ref{pl})  
 but also  Eq.  (\ref{con}).  This means that, with this choice of
$f(R)$,
the existence of  the 
 the SCHW solution, of the RN  solution and  of  the Newtonian limit
for an 
extended, spherically symmetric source is guaranteed. 
Since Eqs.  (\ref{pp1})  and (\ref{c4}) do not hold with the 
choice (\ref {op1}), dressed masses, charges and mass 
densities are different from the bares ones.

With $f(R)$ given by Eq. (\ref{op1}), Eq. (\ref{lamb}) gives,
\begin{equation}\label{lamb2}
 \frac{\Lambda}{ \tilde \Lambda} =  \frac{ 
 e^{16\sigma^{2}\tilde\Lambda^{2} }}{1
+16\sigma^{2}\tilde\Lambda^{2}}.
\end{equation}
Choosing in (\ref{op1}) the dimensional constant $\sigma$ of the same 
order of magnitude of  $1/\tilde\Lambda$:   $\sigma\sim  
4.2/\tilde\Lambda$,  one reproduces the hierarchy of mass scales
(\ref{gf}).

\section{Conclusions}
\lb{SXII}

In this work, we have sought for black hole solutions in an
alternative theory of gravity with an explicit non-minimal coupling
between matter and curvature. General black hole solutions satisfying
the known energy conditions were considered including the ones with
 anti-de Sitter background. Of particular relevance is
the NEC, whose saturation or not, was shown to be a useful criterion
to study general classes of solutions. 

We have shown that  in what concerns the usual  black hole type 
solutions (Schwarzschild and Reissner-Nordstrom), the
NMC  affects very weakly the features of the black hole. 
The analytical form of the solutions remains as those of GR but with 
"bare" GR mass, charge and cosmological constant, are replaced by the
corresponding "dressed" quantities which acquire a contribution from
the non minimal coupling at a relevant curvature scale (c.f. sections
VI, VII and VIII). 

We have fall short of presenting a rigorous prove of the 
uniqueness of the static solutions in NMC theories analogous to 
Birkhoff's theorem in GR. However, we have shown that  under
reasonable assumptions on the analyticity of $f(R)$ and on the
behaviour of the weak 
field limit the black hole solutions we have found in this paper are 
essentially unique.

We have also shown that the existence of black hole solutions as
well as the conditions for a suitable weak field limit constitute a
relevant constraint on the coupling function, $f(R)$, which sets the
strength of the nonminimal coupling between matter and curvature.
These constraints are particularly interesting as they provide  
information both on the local behavior of $f(R)$ near the singularity 
- where quantum gravity effects are expected to dominate - and at
the 
asymptotic region - where gravity is described by the Newtonian weak 
field limit. The knowledge about the coupling function acquired in
this paper, together with other constraints coming from previous
investigations  
on large scale behaviour \cite{DM mim1,DM mim2,DE mim1,review}, 
represent a useful guide for future investigation on theories of 
gravity with NMC between matter and curvature.

Finally, we have shown that the dressing of the  bare
cosmological constant trough the coupling function $f(R)$ can be used
to generate in a natural way a 
hierarchy of  $120$  orders of magnitude between the 
bare and the dressed cosmological constant. Since we are working 
with an effective theory of gravity, this cannot be seen as a full 
solution of the cosmological constant problem. Nevertheless, it 
clearly shows that NMC theories of gravity can mimic 
in a very efficient way exotic forms of energy, most particularly
when  
the solution of the problem requires bridging the short and large 
scale  behaviour.

An important issue that we have not addressed in this paper, which 
deserves future investigation,  is the possibility of existence of a
weak field limit different from the usual one.
Existence of the usual Newtonian  weak field limit has played a 
crucial role in our investigation of black hole solutions in NMC 
theories. Furthermore, in combination with analyticity of the
coupling function 
$f(R)$, it is essential to argue about the uniqueness of the black
hole solutions. 
Moreover, its existence gives local constraints on $f(R=0)$.

The existence of the usual Newtonian limit  is a very 
natural requirement for black holes. This is not necessarily true 
if one considers different regimes of the gravitational interaction, 
where the existence of non-Newtonian weak field limit could
represent an alternative to dark matter \cite{cm,cm1}.

In principle, NMC matter-curvature theories of gravity offer a
suitable framework for 
an alternative scenario in which different regimes of the 
gravitational interaction are described by a different weak field 
limit and by a different local behaviour of the coupling function
$f(R)$  
(c.f. section \ref{ops}).

\begin{acknowledgments}
The work of one of us (O.B.) was partially supported by FCT
(Portugal) under the projects
PTDC/FIS/111362/2009 and CERN/FP/116358/2010.
\end{acknowledgments}

\bigskip

\bigskip

\bigskip

\bigskip

\bigskip

\bigskip

\bigskip

\bigskip

\end{document}